\documentstyle{l-aa}
\input psfig.tex

\def\spose#1{\hbox to 0pt{#1\hss}}
\def\simlt{\mathrel{\spose{\lower 3pt\hbox{$\mathchar"218$}}
     \raise 2.0pt\hbox{$\mathchar"13C$}}}
\def\simgt{\mathrel{\spose{\lower 3pt\hbox{$\mathchar"218$}}
     \raise 2.0pt\hbox{$\mathchar"13E$}}}

\begin{document}

\thesaurus{11.03.5; 11.09.3; 11.05.2; 11.01.1}
\title{Evolution with redshift of the ICM abundances}
\author{Agostino Martinelli$^{1}$, Francesca Matteucci$^{2,3}$ 
and Sergio Colafrancesco$^{4}$}
\offprints{A. Martinelli}
\institute{
Dipartimento di Astronomia, Universita' La Sapienza, Via Lancisi 29, Roma\and
Dipartimento di Astronomia, Universit\`a di Trieste, Via G.B. Tiepolo, 11,
34131 Trieste, Italy \and
SISSA/ISAS, Via Beirut 2-4, 34014 Trieste, Italy \and
Osservatorio Astronomico di Roma, Monte Porzio, Roma, Italy}

\maketitle

\begin{abstract}
We predict the behaviour of the abundances of $\alpha$-elements 
and iron in the
intracluster medium as a function of redshift in poor and rich clusters.
In order to do that we calculate the detailed chemical evolution of 
elliptical galaxies by means of one-zone and multi-zone models and 
then we integrate the 
contributions to the total gas and single elements by ellipticals 
over the cluster mass function at any given cosmic time which is 
then transformed into redshift according to the considered cosmological model.

In the case of the multi-zone model for ellipticals the more external 
regions
evolve much faster than the internal ones which maintain a very 
low level of star formation almost until the present time. In other words, the 
outermost regions develop a galactic wind, after which the region 
evolves passively,
at much earlier times than the innermost regions as opposed to 
the classic one-zone model where the galactic wind develops 
at early times over 
the whole galaxy. We refer to the one-zone model as to {\it burst model} 
and to the multi-zone model as to {\it continuous model}. 
We find that in the case of the burst model the ICM abundances should be
quite constant starting from high redshifts ($z > 2$) up to now, 
whereas in the continuous  model the ICM abundances should increase
up to $z \sim 1$ and are almost constant from $z \sim 1$ up to $z=0$.

Particular attention is devoted to the predictions 
of the $[\alpha/Fe]_{ICM}$ ratio 
in the ICM:
for the 
burst model we predict $[\alpha/Fe]_{ICM}>0$ over the whole range
of redshifts and in particular at $z =0$,  
whereas in the case of the continuous model we predict 
a decreasing $[\alpha/Fe]_{ICM}$ ratio with decreasing $z$
and  $[\alpha/Fe]_{ICM} \le 0$ at $z= 0$. 
In particular, we predict $[O/Fe]_{ICM}(z=0) \le +0.35$ dex
and $[Si/Fe]_{ICM}(z=0) \le +0.15$ dex for the bursts models and 
$[O/Fe]_{ICM}(z=0) \le -0.05$ dex and $[Si/Fe]_{ICM}(z=0) \le +0.13$ dex
for the continuous models, the precise values depending on the assumed 
cosmology.
Finally, we discuss the 
influence of different cosmologies on the results. 

\end{abstract}

\section{Introduction}
The bulk of the observed X-ray emission from 
galaxy clusters is due to thermal bremsstrahlung
in a hot gas ($10^7-10^8 K$) enriched in heavy elements.
In the past two decades a great deal of attention has been devoted to
the study of the abundances of heavy elements (mostly iron) in the
intracluster medium (hereafter ICM).  
From the hot X-ray emitting intergalactic gas in clusters of
galaxies a universal abundance of iron of roughly
1/3 solar has been derived (Renzini 1997 and references therein; 
Fukazawa et al. 1998).
Attempts to explain this ICM iron abundance 
can be traced
back to the early 1970s, when for the first time the iron-emission line
in the X-ray spectra of clusters of galaxies was discovered (Mitchell et al.
1976; Serlemitsos et al. 1977).
Some interpreted the presence of iron in the ICM to be due to gas
ejected from galaxies, either by means of galactic winds or ram pressure
stripping
(Gunn and Gott 1972; Larson and Dinerstein 1975;
Vigroux 1977; De Young 1978; Sarazin 1979; Himmes and Biermann 1980; 
Matteucci and Vettolani 1988; Renzini et al. 1993;
Matteucci and Gibson 1995; Gibson and Matteucci 1997), while
others suggested
pregalactic objects such as population III stars as the origin
(White and Rees 1978).
\par
Recently, thanks to the results obtained with the ASCA satellite,
more detailed observations of iron and $\alpha$-element abundances in
local galaxy clusters are becoming available.
Before the launch of ASCA,
spectroscopic measures of elements such as Si and S were known for
the Perseus cluster (Mushotzky et al. 1981) and A576 (Rothenflug et al.
1984), indicating that the abundances of these elements
as well as that of iron are
roughly solar, although their level of accuracy did
not
allow one to make strong statements about possible differences between the
$\alpha$-elements and iron.
Concerning oxygen there were only a few of measures of the O VIII
line in Virgo and Perseus (Canizares et al. 1988)
indicating a higher
than solar [O/Fe].   
More recent measurements from ASCA
by
Mushotzky et al. (1996) implied $[\alpha/Fe] \approx$ +0.2-+0.3 dex,
thus indicating a general
overabundance of $\alpha$-elements relative to iron in the ICM.
We derived from the Mushotzky data an average
$<(O/Fe)/(O/Fe)_{\odot}> \approx
3.05 \pm 2.19$ (we estimated a formal statistical uncertainty and 
no systematic effect). 
This value  corresponds (by adopting the photospheric 
abundances of Anders and Grevesse (1989)) to 
[O/Fe]$ \approx +0.48^{+0.24}_{-0.55}$ dex.
This means, by considering the errors,
a marginal (if any) overabundance of oxygen relative to iron.
The average Si/Fe ratio 
estimated from Mushotzky data is
$<(Si/Fe)/(Si/Fe)_{\odot}> \approx 2.32 \pm 1.30$ corresponding to
[Si/Fe]= 
$+0.37^{+0.17}_{-0.35}$ dex, which agrees  
with the same ratio obtained by 
Fukazawa et al. (1998) in their analysis of about 40 nearby poor and rich
clusters. 
In fact, from this latter compilation of data we derived
$<(Si/Fe)/(Si/Fe)_{\odot}> \approx 3.15 \pm 1.60$ for clusters with $T > 3$ keV
(we consider these clusters as rich according to our definition in Tab. 2)
and $<(Si/Fe)/(Si/Fe)_{\odot}> \approx 1.39 \pm 0.90$ for clusters with 
$T \le 3$ keV (poor clusters).

Ishimaru and Arimoto (1997a,b)
and Arimoto et al. (1997) claimed that the present uncertainties both on the 
assumed solar abundances (used to derive the quantity [O/Fe]) and 
on the derived 
X-ray abundances are consistent with almost solar values of the 
$[\alpha/Fe]$ ratios in the ICM. 
In particular, the use of the meteoritic abundances for the sun 
(Anders and Grevesse 1989)
instead of the
photospheric ones used in the previous studies, 
would reduce the $[\alpha/Fe]$ values quoted above by $\approx
0.16$ dex (Ishimaru \& Arimoto 1997a,b; Renzini 1977).

Therefore, we consider all of these observational results 
still preliminary and subject to variation.

Concerning the evolution of the abundances and abundance ratios as 
a function of redshift very little is known at the moment. 
However, preliminary results 
by Mushotzky
and Lowenstein (1997) seem to indicate no evolution in the Fe abundance
for $z \simeq 0.5$.

In this paper we study the chemical evolution of the ICM, 
in particular we predict the evolution of Fe 
and $[\alpha/Fe]$ ratios as  
functions of redshift.
This is because the value of the $[\alpha/Fe]$ ratio as well as 
the Fe abundance
in the ICM impose strong
constraints on the evolution of the galaxies in clusters as well as on the 
roles of supernovae of different types in the enrichment of the ICM, 
as discussed by 
Renzini et al. (1993) and Matteucci and Gibson (1995).

Since the heavy elements in the ICM are likely to come from ellipticals, 
as shown by Arnaud (1994), it is important
to explore different kinds of models for the evolution of ellipticals.
In particular, we will use either a one-zone model
(Matteucci and Gibson 1995, 
hereafter referred to as ``burst model'') 
and a multi-zone model 
(Martinelli et al. 1998, hereafter ``continuous model'') 
for elliptical galaxies. 
The continuous model, 
which predicts abundance gradients 
in ellipticals in very good agreement with the observed ones, 
assumes that the outermost regions of these galaxies experience 
early galactic winds whereas the innnermost ones keep an active, 
although low, star formation rate until late times.
The physical reason for this resides in the fact that 
the binding energy of the gas is lower in the outermost galactic regions
and this is supported by the observational finding of a correlation between metallicity and escape velocity in ellipticals (Carollo and Danziger, 1994).
The late occurrence of galactic winds in the most internal regions 
of ellipticals, if true,
will have an effect on the chemical evolution of the ICM,
certainly at variance with the predictions by models with only early winds 
(Matteucci and Gibson 1995; Gibson and Matteucci, 1997).

The plan of this paper is the following.
In Section 2 we present the models and the computational method.
In section 3 the results are discussed and some conclusions are drawn.
We use throughout the paper an adimensional Hubble parameter $h
=H_0/(100~km~s^{-1}~Mpc^{-1})$.

\section{Computational method of the heavy element abundances in galaxy
clusters}

The mass in the form of specific chemical elements as well as the total
gas masses ejected by the elliptical galaxies both at early and late
galactic lifetimes into the ICM, can be computed in detail once a chemical
evolution model
is assumed.

Two different models for the evolution of elliptical
galaxies have been taken into account:

\begin{enumerate}

\item {\it burst models}, namely those with only one major episode of galactic
wind which occurs simultaneously all over the galaxy;

\item {\it continuous models}, where the galactic wind starts at an early
epoch in the most external regions of the galaxies 
and continues until the present time
in the most internal ones.

\end{enumerate}

The burst model is the same as in Matteucci and
Gibson (1995), where we direct the reader for details, in
the case of the Arimoto \& Yoshii (1987) IMF. 
This is their best
model and we refer to it as CWM.

The continuous wind model is described in Martinelli et al. (1998),
where we direct the reader for details. The basic difference of this approach, 
relative to the model of Matteucci and Gibson (1995), is that 
the elliptical galaxy is divided in several shells not interacting. 
For each shell the potential energy of the gas is 
calculated and compared with the thermal energy of the gas  
due to supernova explosions.
A galactic wind develops first in the outermost shells and then progressively in the most inner ones.
An improved
expression for the cooling time of SN (of both
types) remnants as well as adetailed description of 
the potential energy of the gas
are used in this model.
The SN remnant cooling time strongly influences the occurrence of a galactic
wind as well as the amount of energy that SNe can transfer into the interstellar medium (ISM).
In the CWM 
model, a cooling time not dependent on time was used, whereas
in the continuous models the cooling time depends on the gas density
($t_c \sim \frac{1}{\sqrt{\rho_{gas}}}$)
which is a function of time.

In both models, after the occurrence of a galactic wind it is assumed 
that star formation is no longer taking place, and the galaxy 
or the galactic region evolves passively after that.
In the case of the continuous model, the innermost regions (inside 30 pc)
never develop a galactic wind and the star 
formation continues until late times although it is
so
low that it would be hardly detectable.

\begin{figure}[h]
\psfig{figure=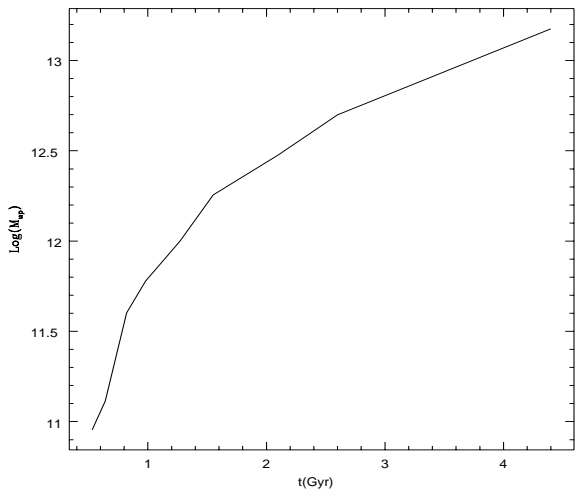,height=7.cm,width=8.cm}
\caption{Mass $M_{up}(t)$ (in solar masses) of the
most massive galaxy which develops a galactic wind at the 
time $t$, in the case of the burst model.}
\end{figure}

The chemical evolution of the ICM is caused by two different processes:
\begin{enumerate}

\item In the burst model, galaxies begin to eject the gas only after the time 
of occurrence of a galactic wind ($t_{GW}$), 
and this process lasts only for few $10^{8}$ years.
The time $t_{GW}$ is a function of the initial galactic mass 
which influences the
potential well.  
Therefore, when the time of the first occurrence of the galactic wind 
increases, namely when more and more massive galaxies achieve the
conditions to develop a wind, the number of
galaxies contributing to the enrichment of the ICM also increases, 
as shown in figure 1.
This type of evolution stops at the time $t_{GW Max}$,
corresponding
to the onset of the galactic wind for the galaxy with the highest mass
considered.

\item In the continuous models, the ICM evolution is caused
by the gas ejected continuously by each galaxy after the time $t_{GW}$.
This continuous ejection process, not present in the burst models, 
has the characteristic
of acting also at low redshift and, in particular, now.
The assumed IMF is the same as in the burst model.

\end{enumerate}

We now elucidate, in some detail, 
how we have calculated the chemical abundances
in the ICM for several values of redshift, according to the two
elliptical galaxy models described above.

\subsection{Burst models}

We computed the ejected masses of each element
(Fe, Si, O, and total gas) as functions of several
initial galactic masses (the procedure we describe here can be also 
found in Matteucci and Vettolani, 1988). We found that
the relation between the generic chemical element $i$ and the initial galactic
mass can be approximated by a power law:

\begin{equation}
M_i=E_iM_{G}^{\beta_i}
\end{equation}

\noindent
where $M_i$ represents the mass ejected in the form of the chemical species
$i$ by a galaxy of initial mass $M_G$, and $E_i$ and $\beta_i$ are two
constants (that can be fixed by a least squares fit).
We have normalized the relations (1) to a galaxy at the {\it break}
of the luminosity function of a cluster (Schechter 1976)
having a blue luminosity $L_*$, absolute blue magnitude $M_*$
and total mass $M_{G_*}$. 

In particular:

\begin{equation}
\frac{M_i}{M_{i_*}}=\left(\frac{M_G}{M_{G_*}}\right)^{\beta_i}=
\left(\frac{L}{L_*}\right)^{\beta_i}
\end{equation}

\noindent
where $M_{i_*}$ represents the mass ejected in the form of an element $i$
by the galaxy with mass $M_{G_*}$.

\begin{table}
\begin{center}
\begin{tabular}{lll}
\hline
$z$ & $\Omega_0=1$, $h=0.5$ &
$\Omega_0=0.4$, $\Omega_{\Lambda}=0.6$, $h=0.6$  \\
\hline
\hline
$0.0$ & $13$ & $13$ \\
\hline
$0.3$ & $8.5$ & $9.4$ \\
\hline
$0.5$ & $6.8$ & $7.6$ \\
\hline
$1.0$ & $4.4$ & $5.0$ \\
\hline
$2.0$ & $2.1$ & $2.6$ \\
\hline
$3.0$ & $1.3$ & $1.6$ \\
\hline
$4.0$ & $0.82$ & $0.98$ \\
\hline
$5.0$ & $0.53$ & $0.64$ \\
\hline
\hline
\end{tabular}
\caption{Age of the elliptical galaxies for several values of
the redshift $z$ in the two cosmological models examined}
\end{center}
\end{table}

For each redshift $z$ we evaluated, for the cosmological model considered,
the corresponding age $t$ of the elliptical galaxies (Table 1).
Then, for each $t$
we determined from figure 1 the mass $M_{up}(t)$ of the
most massive galaxy which develops a galactic wind
just at the time $t$.
The maximum mass considered is $M_{up}(t)=M_{up}=2 \cdot 10^{12}M_{\odot}$.
Let $M_{low}$ ($10^{8} M_{\odot}$)
be the smallest mass for elliptical galaxies in our models.
Let finally $L_{up}(t)$ and $L_{low}$ be the luminosities related to
the previous mass bounds.
Therefore by integrating over all the galaxies with luminosity
between $L_{low}$ and $L_{up}(t)$, we show that the total mass in the form
of an element $i$, ejected into the ICM is:


\begin{equation}
M_{i}^T=\int_{L_{low}}^{L_{up}(t)}\Phi(L/L_*)(L/L_*)^{\beta_i}d(L/L_*)
\end{equation}

\noindent
where $\Phi(L/L_*)(L/L_*)^{\beta_i}d(L/L_*)$ represents the mass of an element
$i$ per interval of $d(L/L_*)$. The masses computed
by eq. (3) are therefore expressed in units of an element $i$
ejected by a galaxy of luminosity $L_*$, namely of mass $M_{G_*}$ 
(the luminosity
and the mass at the break of the luminosity function).
The luminosity function $\Phi(L/L_*)$ is taken to be 
Schechter (1976), namely:

\begin{equation}
\Phi(L/L_*)=n^*(L/L_*)^{\alpha}exp(-L/L_*)
\end{equation}

\noindent
where $n^*$ is a measure of the cluster richness and $\alpha =1.3$. 

Integrating the previous
eq. (3) one obtains:

\begin{equation}
M_{i}^T=n^* f [\Gamma(\alpha+1+\beta_i,L_{low}/L_*)-
\Gamma(\alpha+1+\beta_i,L_{up}(t)/L_*)]
\end{equation}

\noindent
where $\Gamma(a,b)$ is the incomplete $\Gamma$ function and $f$ represents
the fraction by number of ellipticals in a cluster.
The parameters $f$, $n_*$, $\alpha$ and $M_*$ are fixed for each
individual cluster.
We consider here the evolution of the ICM chemical abundances for two
galactic systems: a rich cluster (RC) and a poor cluster (PC).
In Table 2 the values of the parameters related to these 
prototype clusters are given.

\begin{table}
\begin{center}
\begin{tabular}{lllll}
\hline
Cluster & $\alpha$ & $f$ & $n_*$ & $M_*$ \\
\hline
\hline
RC & $-1.3$ & $0.8$ & $115$ & $-22.5$ \\
\hline
PC & $-1.3$ & $0.3$ & $20$ & $-22.0$ \\
\hline
\hline
\end{tabular}
\caption{Input ingredients for rich and poor cluster models.}
\end{center}
\end{table}

At this point we need to express eq. (5) as a function of the 
galactic mass instead of the luminosity. We obtain:

\begin{eqnarray}
M_{i}^T &=& E_i n^* f (h^2K)^{\beta_i}10^{-0.4\beta_i(M_*-5.48)}\\
\nonumber
~&~& [\Gamma(\alpha+1+\beta_i,(M_{low}h^2/K)10^{-0.4(M_*-5.48)})-\\
\nonumber
~&~& \Gamma(\alpha+1+\beta_i,(M_{up}(t)h^2/K)10^{-0.4(M_*-5.48)})]
\end{eqnarray}


\noindent
where $K=M_G/L$ is the mass to luminosity ratio,
with $M_G$ expressed in solar masses and $L$
in solar luminosities, $h^{-1}=H_0/100$ with $H_0$ the Hubble constant.
The mass to luminosity ratio is computed as a function of the assumed IMF,
as described in Matteucci and Gibson (1995).

\subsection{Continuous models}
In the case of continuous models
the computation is different from the previous case because now the quantities
$E_i$ and $\beta_i$ are time dependent.
For each value of $t$ we calculate the mass of gas in the form of
a specific chemical element
ejected from each galaxy.
Again we can find the corresponding values of
$E_i(t)$ and $\beta_i(t)$ by using a least squares fit.
By substituting these quantities in eq. (6) we then can calculate
the total mass of the chemical element $i$
ejected into  the ICM at the time $t$.

\vskip 0.8cm

\begin{table}
\begin{center}
\begin{tabular}{lll}
\hline
Mod. & Cluster & Cosmology \\
\hline
\hline
1  & RC & $\Omega_0=1$, $h=0.5$ \\
\hline
2  & RC & $\Omega_0=0.4$, $\Omega_{\Lambda}=0.6$, $h=0.6$ \\ 
\hline
3  & PC & $\Omega_0=1$, $h=0.5$ \\ 
\hline
4  & PC & $\Omega_0=0.4$, $\Omega_{\Lambda}=0.6$, $h=0.6$ \\ 
\hline
\hline
\end{tabular}
\caption{Models considered.}
\end{center}
\end{table}

The evolution of the elliptical galaxies is considered in the framework
of different cosmological models:
a flat, scale invariant CDM cosmology with $\Omega_0=1$, $h=0.5$
and a vacuum dominated CDM model with 
$\Omega_0=0.4$, $\Omega_{\Lambda}=0.6$ and  $h=0.6$.

In all the models considered we set the galaxy formation
epoch, $z_f=10$, and their age $=13$ Gyr (we consider the same age for
all the galaxies in our code).

The models are defined in Table 3.

\begin{figure}[h]
\psfig{figure=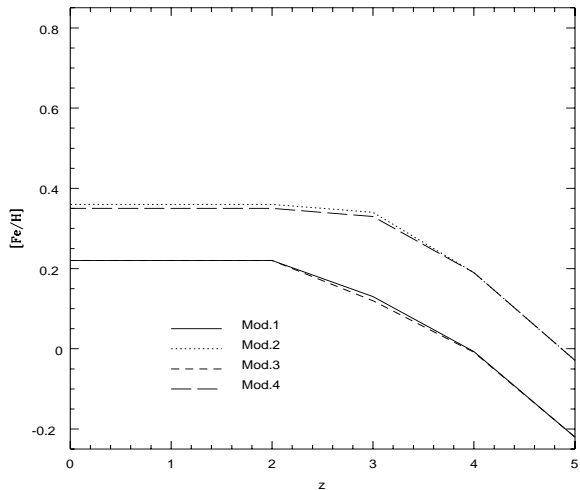,height=7.cm,width=8.cm}
\caption{$\left[\frac{Fe}{H}\right]=log(Fe/H)-log(Fe/H)_{\odot}$ 
in the total gas ejected from
the ellipticals as a function of the redshift
in the case of burst models.The curves refer to a poor and a rich clusters and
to different cosmologies as shown in Table 3. The adopted solar abundances are the meteoritic abundances from Anders and Grevesse (1989).}
\end{figure}

\begin{figure}[h]
\psfig{figure=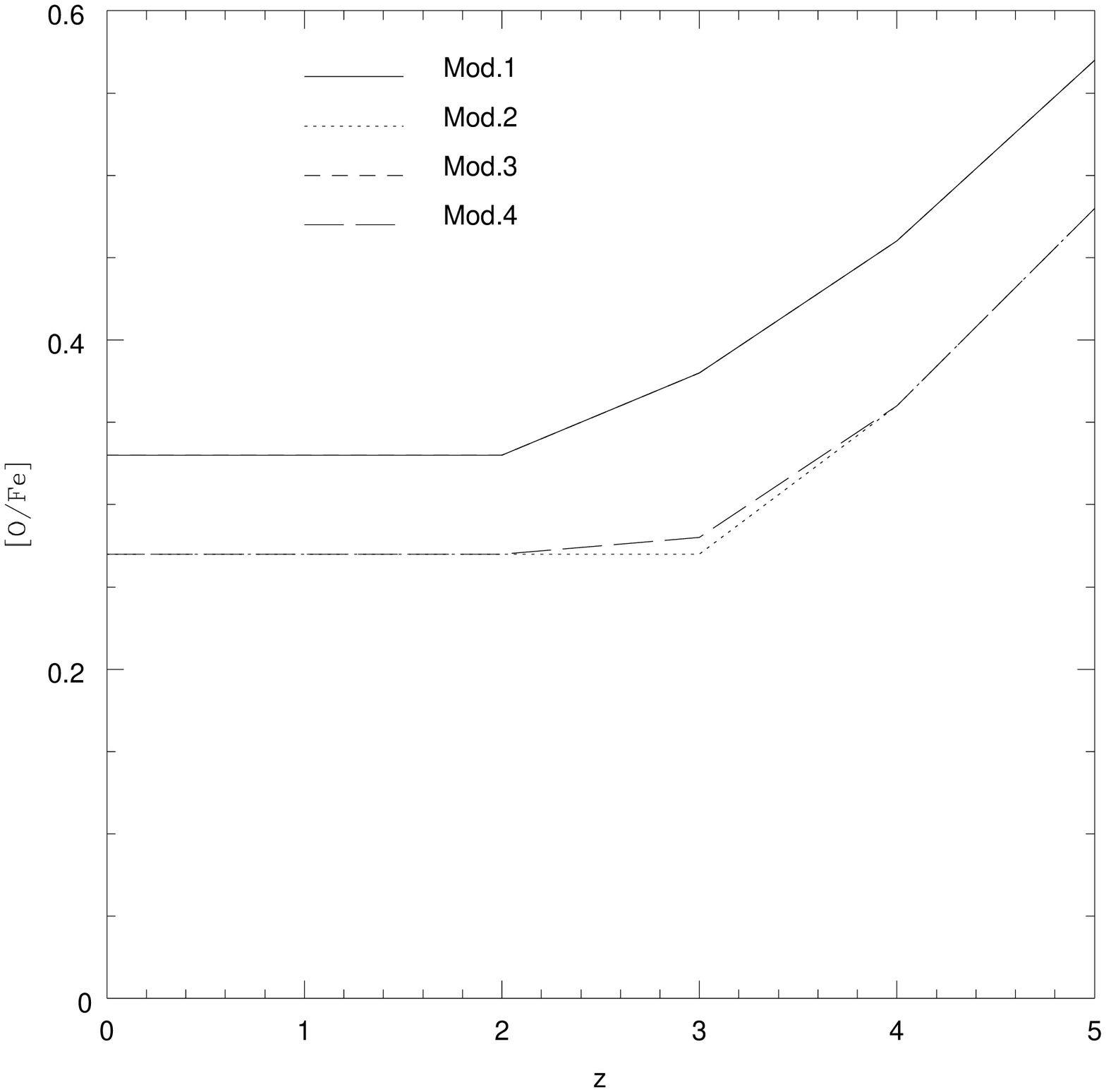,height=7.cm,width=8.cm}
\caption{$\left[\frac{O}{Fe}\right]$ in the total gas ejected from
ellipticals and in the ICM as a function of the redshift
in the case of burst models. The curves refer to a poor and a rich 
cluster and different cosmologies as shown in Table 3.}
\end{figure}

\section{Results and Conclusions}

\begin{figure}[h]
\psfig{figure=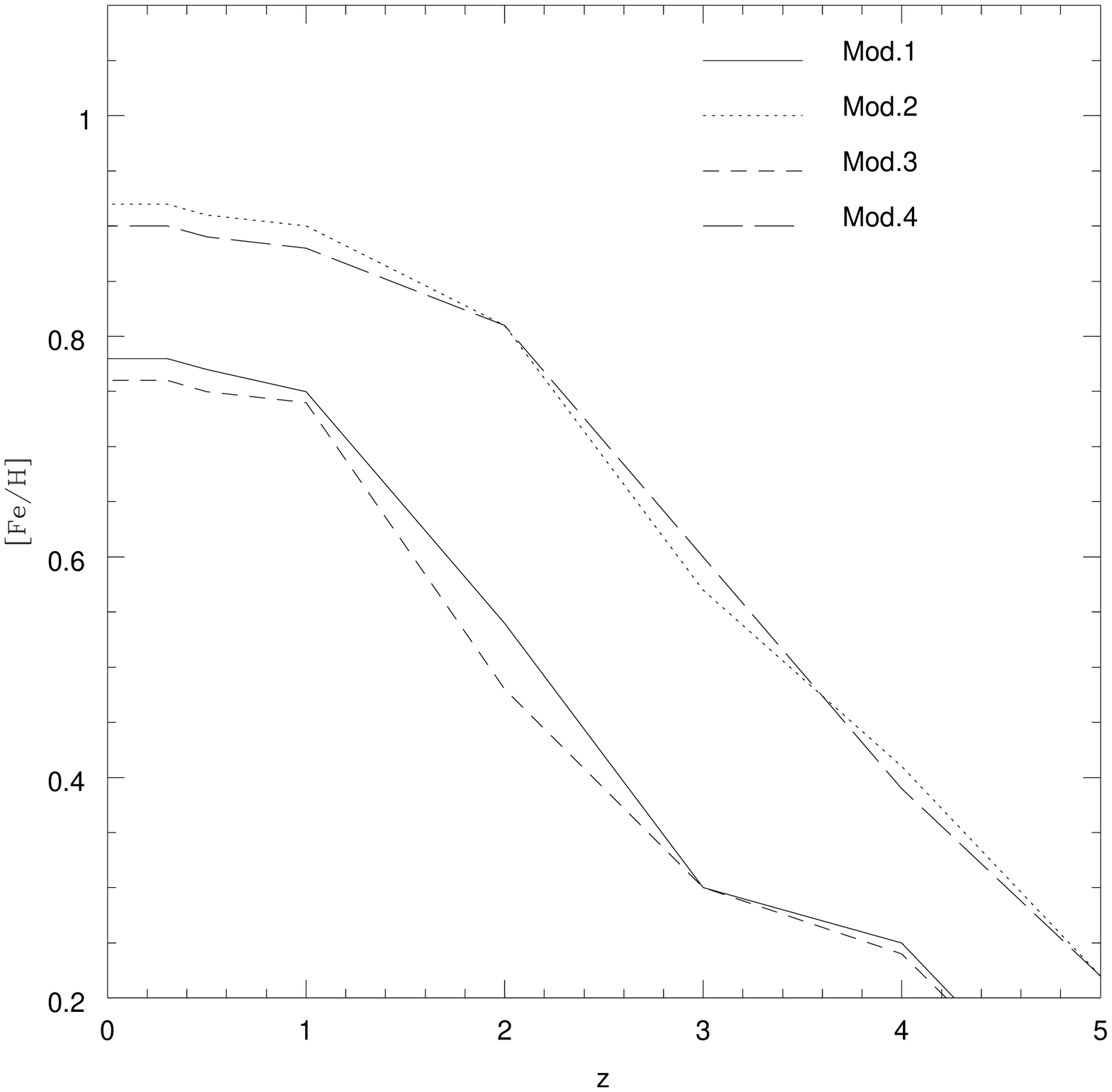,height=7.cm,width=8.cm}
\caption{$\left[\frac{Fe}{H}\right]$ in the total gas ejected from 
ellipticals and in the ICM as a function of the redshift
in the case of continuous models.}
\end{figure}

In this paper we have computed the evolution of the abundances of 
Fe and $\alpha$-elements in the ICM of poor and rich clusters 
as a function of redshift under different cosmologies. 
To do that we have used two different models for the chemical evolution 
of elliptical galaxies, which are the main 
contributors to the ICM enrichment.
In particular, we used a model where the galactic wind develops 
at the same time all over the galaxy (the burst model) and
after that the galaxy evolves only passively not contributing 
any longer to the ICM enrichment, and a model where galactic 
winds develop first at large galactocentric distances 
and then progressively later in the innermost zones (continuous model).

Our results are shown in 
Figures 2-7, where the behaviour of the
$\left[\frac{Fe}{H}\right]$, $\left[\frac{O}{Fe}\right]$ and
$\left[\frac{Si}{Fe}\right]$
as functions of the redshift $z$ in the cases of
burst models (Fig. 2, 3, 6) and continuous models (Fig. 4, 5, 7) is
indicated.
It should be noted that the [Fe/H] and the [$\alpha$/H] abundances 
refer to the abundances expected in the total gas ejected by 
ellipticals and not mixed with the pristine gas in the cluster.
Previous calculations (Matteucci and Vettolani, 1988; 
Matteucci and Gibson, 1995;
Gibson and Matteucci 1997) have shown that while the predicted total iron mass
ejected by ellipticals agrees with the observed iron mass in clusters 
the total mass of gas ejected from ellipticals is much lower than the 
observed total gas mass in clusters. 
This is interpreted as due 
to the fact that most of the gas in clusters should be pristine gas.

The models presented here also predict a too 
small amount of ejected total gas as compared to the observed one. 
A larger production of total gas from galaxies can be obtained 
by adopting a steep faint end slope of the luminosity function. However, 
Gibson and Matteucci (1997) showed that in this case the contribution 
of the dwarf ellipticals is at most $15\%$ of the total ICM gas, so that the conclusion about most of the ICM gas being pristine gas remains unaffected. 
In order to obtain the abundances in the ICM 
we should rescale the abundances shown in figures 2 and 4 
to the total observed 
mass of gas in a typical poor and a rich cluster by assuming 
that most of it is pristine gas.
In order to compute the evolution of the abundances 
as a function of redshift we also assumed that the amount of pristine gas 
in the clusters did not change during the cluster lifetime, 
in agreement with the
results of White et al. (1993) which found that the baryon 
fraction in clusters should have been constant.

After doing that by taking $M_{gasICM_{R}}= 4.4 \cdot 10^{14}M_{\odot}$ as 
representative of the total gas of a rich cluster (Coma) and 
$M_{gasICM_{P}}=2 \cdot 10^{13}M_{\odot}$
as representative of a poor cluster (Virgo) we found that, 
in order to represent  the $[Fe/H]_{ICM}$, the curves 
of figure 2 should be lowered by -1.48 and -1.35 dex for a rich and a poor cluster, respectively, whereas the curves of figure 4 should be lowered by 
-1.37 and -1.30 for a rich and a poor cluster, respectively.
The present time iron abundance, $Fe_{ICM}$, obtained in this way from the
continuous models is in 
very good agreement with the observational estimates of the iron abundance in local clusters. In fact, the continuous models predict 
$Fe_{ICM}(z=0) \sim 0.3 Fe_{\odot}$ both for poor and rich clusters, 
whereas the burst model predicts a lower abundance ($Fe_{ICM}(z=0)
\sim 0.1 Fe_{\odot}$.

\begin{figure}[h]
\psfig{figure=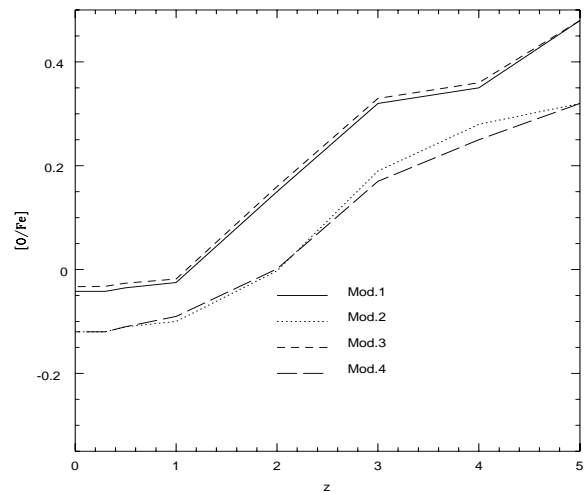,height=7.cm,width=8.cm}
\caption{$\left[\frac{O}{Fe}\right]$ in the total gas ejected from ellipticals
as a function of the redshift
in the case of continuous models}
\end{figure}

\begin{figure}[h]
\psfig{figure=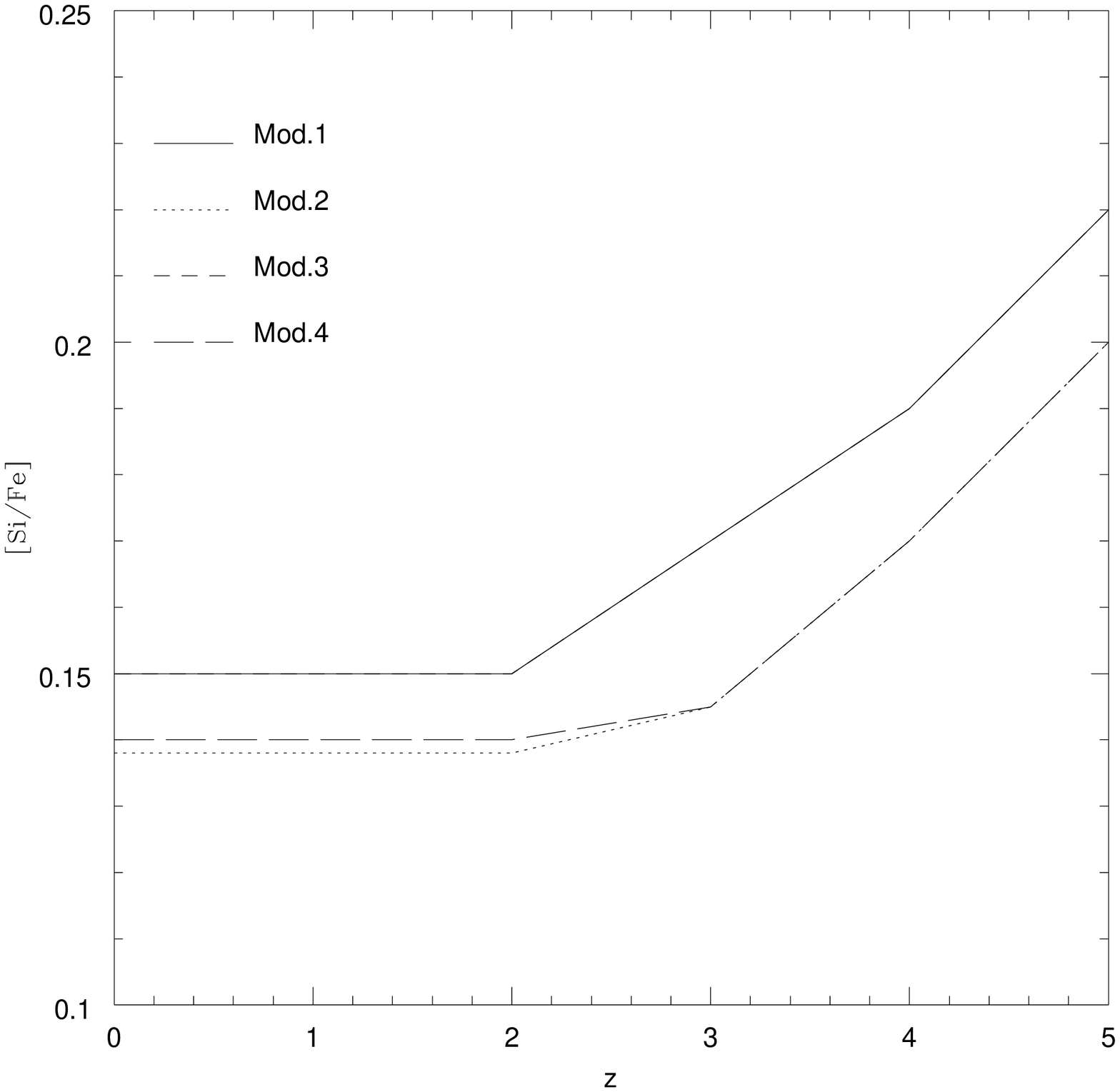,height=7.cm,width=8.cm}
\caption{$\left[\frac{Si}{Fe}\right]$ in the total gas ejected from
ellipticals and in the ICM as a function of the redshift
in the case of burst models. The curves refer to a poor and a rich 
cluster and different cosmologies as shown in Table 3.}
\end{figure}

\begin{figure}[h]
\psfig{figure=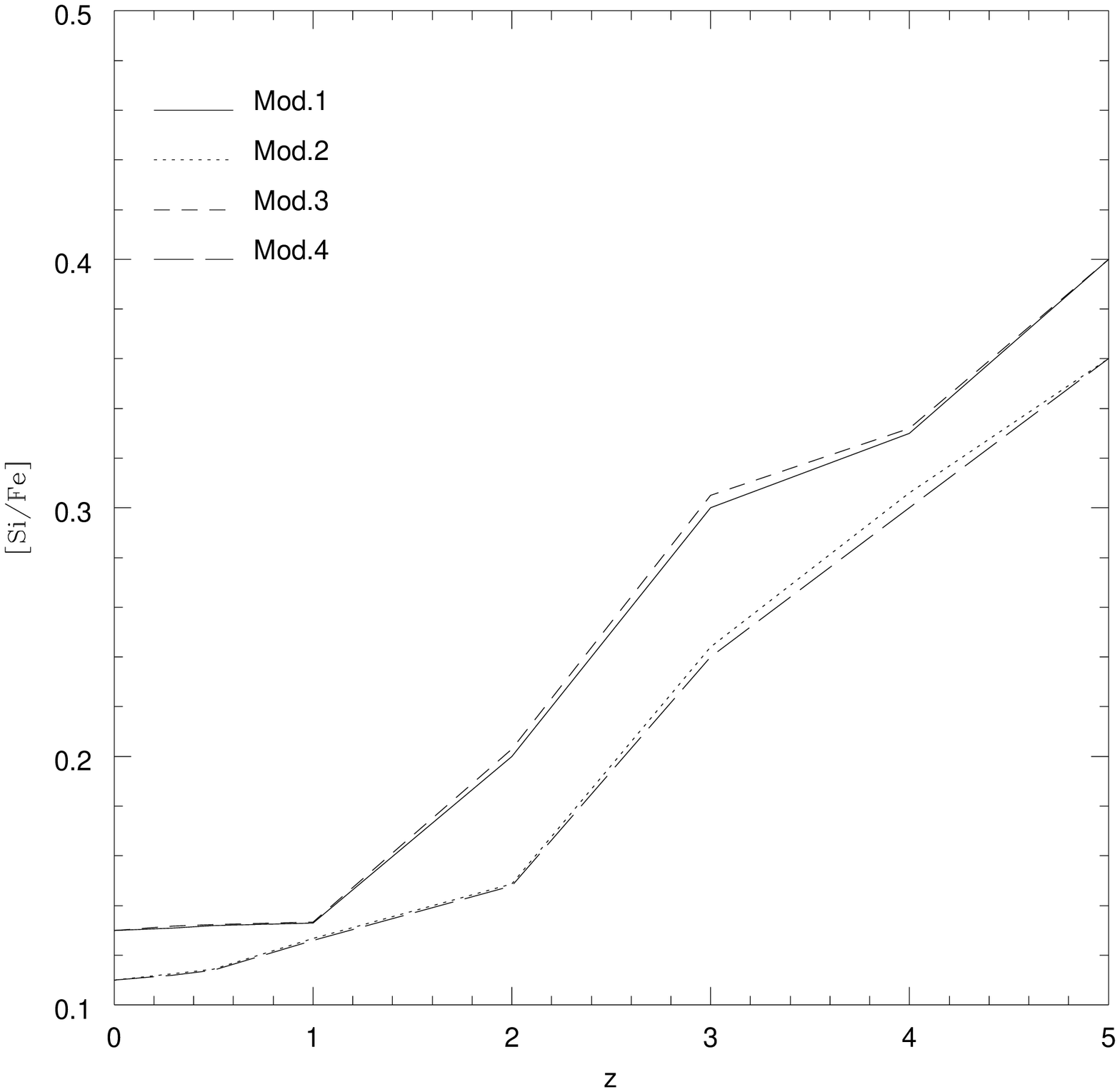,height=7.cm,width=8.cm}
\caption{$\left[\frac{Si}{Fe}\right]$ in the total gas ejected from
ellipticals and in the ICM as a function of the redshift
in the case of burst models. The curves refer to a poor and a rich 
cluster and different cosmologies as shown in Table 3.}
\end{figure}

On the other hand, the abundance ratios [$\alpha$/Fe] are completely 
unaffected by the amount of gas in the clusters and therefore the abundance ratios in the gas ejected from ellipticals represent also the abundance ratios in the ICM.
For this reason we want to stress 
the importance of considering abundance ratios rather than absolute abundances.

From figures 2-7 we can infer the following conclusions:

\begin{itemize}

\item We observe that in the burst models the chemical evolution of the ICM
stops entirely at $z=3$ 
when the cosmological parameters are 
$\Omega_0=0.4$, $\Omega_{\Lambda}=0.6$, $h=0.6$ and at $z=2$ when
$\Omega_0=1$, $\Omega_{\Lambda}=0.0$ and  $h=0.5$, respectively.

\item In the case of continuous models there is a continuous evolution up to
$z\simeq 1$ and negligible evolution from $z \simeq 1$ up to $z=0$.

\item The predicted [Fe/H] at the present time in the ICM is in very good 
agreement with the observations ( 1/3 solar)
for the continuous models whereas the 
burst models 
predict a lower abundance.

\item For the $[\alpha/Fe]_{ICM}$ ratios we predict 
$[\alpha/Fe]_{ICM}>0$ 
at $z=0$ in the burst models and $[\alpha/Fe]_{ICM} \le 0$
at $z=0$ for the continuous models, the exact values 
of the $[\alpha/Fe]_{ICM}$
ratio depending on the assumed cosmology.
In particular, at $z = 0$ 
we predict $[O/Fe]_{ICM} \simeq +0.27 \div +0.37 $ dex
and $[Si/Fe]_{ICM} \simeq +0.14 \div +0.15$ dex for the bursts 
models and 
$[O/Fe]_{ICM} \simeq -0.12 \div -0.05$ dex and 
$[Si/Fe]_{ICM} \simeq +0.11 \div +0.13$ dex
for the continuous models.
These values were derived by adopting the meteoritic solar 
abundances (Anders and Grevesse, 1989). 

In both cases (different normalization to the solar abundances)
the predicted $[\alpha/Fe]_{ICM}$ ratios at the present time are higher in the
burst than in the continuous model.
The main reason for this behaviour of the $[\alpha/Fe]_{ICM}$ 
ratio in the two different cases is the fact that in the burst model
the galactic wind develops at early times in all galaxies and therefore 
it is mainly enriched by the products of SNe II (high $[\alpha/Fe]$ ratio).
On the other hand, in the continuous models the galactic wind continues 
to develop until the present time and therefore it is enriched also by
the products of SNe Ia (low $[\alpha/Fe]$ ratio). 
This result is quite important since it depends uniquely on the evolution 
of the ellipticals and not on the particular history of the cluster 
evolution
(infall, etc..) as it is for absolute abundances.
We note that the predicted present time  $[Si/Fe]_{ICM}$ ratios 
are larger than the present time $[O/Fe]_{ICM}$ for the continuous models.
This is due to the different nucleosynthesis of Si and O: 
O, in fact, is entirely produced by SNe II whereas Si is mostly 
produced by SNe II but a fraction, not entirely negligible, arises from 
SNe Ia and this contributes to avoid that the Si/Fe
ratio decreases too much, since the bulk of Fe coming from SNe Ia
is partly compensated by the Si which comes also from SNe Ia.
For this reason Si is not the best $\alpha$-element to be used to check the 
$[\alpha/Fe]$ ratio, whereas Mg and O should be preferred.

Finally, it is worth noting that these $[\alpha/Fe]_{ICM}$ ratios have 
been obtined by adopting a flat IMF in the galactic models ($x=0.95$). 
These values are therefore upper values especially in the case 
of the continuous model,
since the adoption of the Salpeter IMF ($x=1.35$) would lead to lower
present time values
of these ratios, due to the larger contribution from type Ia SNe relative to type II SNe.

\item For both galactic models these results do not differ
for rich or poor clusters. 

\item The effect of a change in the cosmological parameters 
is a simple translation
of the abundance ratios (namely the $[\alpha/Fe]$ ratios
are lower by at most $\sim 0.1$ dex
when $\Omega_0=0.4$ than when $\Omega_0=1$). A change in 
the cosmology produces 
also a delay in the chemical evolution, in the sense that
when $\Omega_0=1$ the chemical evolution of the ICM
stops a little later than when $\Omega_0=0.4$.
This is a consequence of the different correspondence
between the redshift $z$ and the galactic age 
$t$ as shown in Table 1.

\end{itemize}

From the previous results we can conclude that if we 
want to see an evolution
in the ICM we have to be able to perform  observations of abundances
in very high redshift clusters (at least $z \ge 1$).
Preliminary
results from ASCA (Mushotsky and Lowenstein, 1997) seem to indicate 
no evolution in
the abundance of Fe for $z \le 0.5$ in agreement with our results related to
all models here considered. Therefore, on this basis alone 
we cannot distinguish among the two models.
Concerning our predictions on the $[\alpha/Fe]_{ICM}$ ratios, recent studies 
(Ishimaru and Arimoto, 1997a,b) adopting the meteoritic solar abundances of Anders and Grevesse (1989) as in this work,
seem to favor a low value of this ratio in very good agreement with 
our continuous wind model predictions.

\acknowledgements{A. M. would like to thank SISSA for the 
kind ospitality during the development of this work.}

{}


\begin{thebibliography}{}
\bibitem []{} Anders, E., Grevesse, N., 1989 Geochim.Cosmochim. Acta 53, 197
\bibitem[]{} Arimoto N., Yoshi Y., 1987, A\&A, 173, 23
\bibitem[]{} Arimoto N., Matsushita, K., Ishimaru, Y., Ohashi, T., Renzini, A.,
1997, Ap.J., 477, 128
\bibitem[]{} Arnaud, M., 1994, in ``Clusters of Galaxies'', 
eds. F. Durret et al.,
Editions Frontieres, Gif-sur-Yvette, p.211   
\bibitem[]{} Canizares, C.R., Markert, T.H., Donahue, M.E., 1988, {\it Cooling
Folws in Clusters and Galaxies}, ed. A.C. Fabian, Kluwer: Dordrecht p.63
\bibitem[]{} Carollo, M., Danziger, I.J., 1994, MNRAS, 270, 523
\bibitem[]{} De Young, D.S., 1978, ApJ, 223, 47
\bibitem[]{} Fukazawa,Y., Makishima,K., Tamura, T. et al., 1998, PASJ, 50,
187
\bibitem[]{} Gibson B.K., Matteucci, F., 1997, ApJ, 457, 47
\bibitem[]{} Gunn, J.E. and Gott, J.R., 1972, Ap.J., 176, 1
\bibitem[]{} Himmes, A. and Biermann, P., 1980, A\&A, 86, 11
\bibitem[]{} Ishimaru, Y. and Arimoto, N., 1997a, PASJ, 49, 11
\bibitem[]{} Ishimaru, Y. and Arimoto, N., 1997b, IAUS 187, {\it Cosmic
Chemical Evolution}, Kyoto, Japan
\bibitem[]{} Larson R. B. and Dinerstein, H.L., 1975, PASP, 87, 911
\bibitem[]{} Martinelli A., Matteucci F. and Colafrancesco S.,1998
MNRAS 298, 42.
\bibitem[]{} Matteucci, F. and Vettolani, G., 1988, A\&A, 202, 21
\bibitem[]{} Matteucci F. and Gibson B., 1995, A\&A,  304, 11
\bibitem[]{} Mitchell, R.J., Culhane, J.L., Davison, P.J.N., Ives, J.C., 1976,
MNRAS, 176, 29
\bibitem[]{} Mushotzky R.F., Holt, S.S., Smith, B.W., Boldt, E.A.,
1981, ApJ, 225, 21
\bibitem[]{} Mushotzky R.F., Loewenstein, M., Arnaud, M., Tamura, T.
Fukazawa, Y. et al. 1996, Ap.J., 466, 686 
\bibitem[]{} Mushotzky R.F., Lowenstein M., 1997, ApJ, 481, L63
\bibitem[]{} Renzini, A., Ciotti, L., D'Ercole, A., Pellegrini, S., 1993, 
Ap.J.,
419, 52
\bibitem[]{} Renzini, A., 1997, Ap.J., 488, 35
\bibitem[]{} Rothenflug, R., Vigroux, L., Mushotsky R.F., Holt, S.S.,
1984, Ap.J., 279, 53
\bibitem[]{} Schechter, P, 1976, Ap.J., 203, 297
\bibitem[]{} Selermitsos, P.J., Smith, B.W., Boldt, E.A., Holt, S.S.,
Swank, J.H., 1977, Ap.J., 211, L63
\bibitem[]{} Sarazin, C.L., 1979, Astrophys. Lett., 20, 93
\bibitem[]{} Vigroux, L., 1977, A\&A, 56, 473
\bibitem[]{} White, S.D.M. and Rees, M.J., 1978, MNRAS, 183, 341
\bibitem[]{} White, S.D.M., Navarro, J.F., Evrard, A.E., Frenk, C.S. 1993,
Nature, 366, 429






\end{thebibliography}
\end{document}